\documentclass{JHEP}
\usepackage{amsmath,amssymb,amscd,amsfonts,mathrsfs,cite}
\usepackage{bm}
\def\no{\nonumber \\}

\def\a{{\alpha}}

\def\b{{\beta}}
\def\g{{\gamma}}
\def\G{{\Gamma}}
\def\d{\delta}

\def\ep{{\epsilon}}

\def\m{{\mu}}
\def\n{{\nu}}

\def\t{{\theta}}

\def\f#1#2{{\frac{#1}{#2}}}

\def\s{\sqrt}

\def\f {\frac}
\def\ti{\tilde}

\def\p{\partial}

\newcommand{\half}{{1\over 2}}

\def\eps{{\epsilon}}

\title{Central Charges in Extreme Black Hole/CFT Correspondence}

\author{Geoffrey Comp\`ere \\ {Department of Physics, University of California at Santa Barbara, Santa Barbara, CA 93106, USA \\ E-mail: \email{gcompere@physics.ucsb.edu}}}

\author{Keiju Murata \\ Department of Physics, Kyoto University, Kyoto 606-8502, Japan \\ Email:
\email{murata@tap.scphys.kyoto-u.ac.jp}}

\author{ Tatsuma Nishioka \\ Department of Physics, Kyoto University, Kyoto 606-8502, Japan \\ Email:
\email{nishioka@gauge.scphys.kyoto-u.ac.jp}}

\date{\today}

\abstract{The Kerr/CFT correspondence has been recently broadened to the general extremal black holes under the assumption that the central charges from the non-gravitational fields vanish. To confirm this proposal, we derive the expression of the conserved charges in the Einstein-Maxwell-scalar theory with topological terms in four and five dimensions and check that the above assumption was correct. Combining the computed central charge with the expected form of the temperature, the Bekenstein-Hawking entropy of the general extremal black holes in four and five dimensions can be reproduced by using the Cardy formula.
\bigskip\bigskip}

\preprint{KUNS-2187\ arXiv:0902.1001}


\begin{document}

\tableofcontents

\section{Introduction}
Recently, a new duality called the Kerr/CFT correspondence was proposed in \cite{GHSS}. It was shown that
the black hole entropy of the four-dimensional extremal Kerr black hole with angular momentum $J$ 
can be reproduced by the statistical entropy of a dual two-dimensional CFT with the central charge $c=12J$, 
which is evaluated following the approach originally taken by Brown-Henneaux for AdS$_3$ \cite{BrHe}. Such a duality has been generalized to other black holes including higher dimensions
\cite{HMNS,LMP,AOT,Nakayama,CCLP,ITW,AOT2,PW,CW,CJR,LS,Gh,LMPV}\footnote{Note that the appearance of one copy of a Virasoro algebra in the near horizon region of a generic non-extreme black hole whose central charge is proportional to the horizon area was noted earlier in  \cite{Solodukhin:1998tc,Carlip:1998wz}.}.
In these works, the central charge is usually computed only from the gravitational field while 
contributions from other fields like vector and scalar
fields are neglected. Nevertheless, the correct Bekenstein-Hawking
entropy can be reproduced from the Cardy formula.
Thus, we can conjecture that the central charge of extremal black holes comes
from only gravitational part and the contribution from other fields
vanishes. This conjecture was proven for the Kerr-Newman-AdS-dS black hole in the
Einstein-Maxwell theory with cosmological constant \cite{HMNS}.
In this paper, following the cohomological methods
\cite{BaCo0,BaCo2,Compere}, 
we derive the expression of the conserved charges for the
fairly general action,
\begin{equation}
	S = \f{1}{16\pi}\int d^Dx\sqrt{-g}\bigg( R - \frac{1}{2}f_{AB}(\chi)
	\partial_\mu \chi^A \partial^\mu \chi^B - V(\chi)
	- \frac{1}{4}k_{IJ}(\chi)F^I_{\mu\nu}F^{J\mu\nu} \bigg)+S_{top}\
	,
\label{generalaction}
\end{equation}
where $D=4$ or $5$. The ``topological'' term $S_{top}$ is given by
\begin{align}
 &S_{top}= \f{1}{16\pi}\int
 d^4x\s{-g}\,\frac{1}{4}h_{IJ}(\chi)\epsilon^{\mu\nu\rho\sigma}F^I_{\mu\nu}F^J_{\rho\sigma}
\qquad (D=4)\ ,
\\
 &S_{top}= \f{1}{16\pi}\int d^5x\s{-g}\,\frac{1}{2}C_{IJK}
	\epsilon^{\a\b\g\rho\sigma}
	A^I_\a F^J_{\b\g}F^K_{\rho\sigma}
\qquad (D=5)\ .
\end{align}
Then, we evaluate the central charge for general extremal black holes in these theories
and prove the above conjecture.
Combining the result and the expected form of the temperature, 
we reproduce the Bekenstein-Hawking entropy of 
the general extremal black holes by the Cardy formula in four and five
dimensions respectively.
This result supports the extremal black hole/CFT correspondence suggested
in~\cite{HMNS}.


The organization of this paper is as follows.
In section \ref{NH}, we consider 
the near horizon solution of the extremal black hole in the
theory~(\ref{generalaction}). We recapitulate the boundary conditions for the
fluctuations of the near horizon geometry and the Virasoro asymptotic symmetry 
algebra.
In section \ref{Fomalism}, we review the formalism to obtain the
conserved charges for extremal black holes
following~\cite{BaCo0,BaCo2,Compere,Wald,IyWa,Tachikawa}.
In section \ref{4Dcharge}, we obtain the explicit 
expression of the conserved charges of the four-dimensional
Lagrangian~(\ref{generalaction}).
In section \ref{CC4D}, we calculate the central charge of the
Virasoro algebra and find that there is no contribution from
the non-gravitational part. Using this fact, we show that the
Bekenstein-Hawking entropy can be reproduced by the Cardy formula.
In section \ref{CC5D}, we repeat the calculation in the five-dimensional theory~(\ref{generalaction}).
We find that there is no contribution from the non-gravitational 
part and that the Bekenstein-Hawking entropy can be reproduced by the Cardy formula again. 
We conclude in section \ref{Conc}.

\section{Near Horizon Geometry of Extreme Black Holes and Virasoro Algebra}\label{NH}

\subsection{Near horizon geometry of extreme black holes}\label{NHofEB}
We focus on the case that $f_{AB}(\chi)$ and $k_{IJ}(\chi)$ are positive definite and the scalar potential $V(\chi)$ is non-positive in (\ref{generalaction}). It was shown in \cite{KLR,KuLu,KuLu2} that if we assume $D-3$ rotational symmetries with a fixed
point in the asymptotic region and that the horizon topology is not $T^{D-2}$, then 
the near horizon solution of a stationary, extremal black hole solution in 
the general action~(\ref{generalaction}) is given by 
\begin{equation}
\begin{split}
&ds^2 = \Gamma(\theta)\left[-r^2dt^2+\frac{dr^2}{r^2}
+ \alpha (\t) d\theta^2 \right] +
 \sum_{i,j=1}^\ell
\gamma_{ij}(\theta)(d\phi^i + k^irdt)(d\phi^j + k^jrdt)\
 , \\
&\hspace{2cm}
\chi^A = \chi^A (\t), \qquad A^I = \sum_{i=1}^\ell f_i^I(\t) (d\phi^i + k^irdt) \ . 
\end{split}
\label{GenExt}
\end{equation}
where $\ell=1,2$ for $D=4,5$, respectively. Notice that the above metric for $D=5$ can be obtained from the extremal black holes with 
a topologically $S^1\times S^2$ horizon and with a topologically $S^3$ horizon \cite{KLR}.
The near horizon geometry has the enhanced isometry $SL(2,R)\times U(1)^\ell$, as was observed earlier using the attractor mechanism \cite{Astefanesei:2006dd,Astefanesei:2007bf}, and the scalar and vector fields are also invariant under this symmetry.
The horizon of the extremal black hole 
was located at $r=0$.
Thus, the Bekenstein-Hawking entropy is given by
\begin{equation}
 S_{grav}=\frac{(2\pi)^\ell}{4}\int^\pi_0d\theta\sqrt{\Gamma (\theta) \alpha (\theta)
  \gamma (\theta)}\ ,
\label{Sg}
\end{equation}
where we denote $\gamma(\theta) \equiv \text{det}(\gamma_{ij}(\theta))$.
We will consider the dual CFT description of~(\ref{GenExt}) and
reproduce the Bekenstein-Hawking entropy~(\ref{Sg}).

\subsection{Virasoro algebra in four dimensions}\label{V4D}
Now, we consider fluctuations of the near horizon geometry
of the extremal black hole~(\ref{GenExt}).
We should specify the boundary conditions for the fluctuations at
$r=\infty$. We adopt the boundary conditions given in~\cite{GHSS,HMNS}, 
which are determined in order to obtain the Virasoro algebra as the asymptotic
symmetry group.
In four dimensions, the boundary conditions are given by
\begin{equation}\label{metricbc}
h_{\mu\nu} \sim \mathcal{O}
\left(
  \begin{array}{cccc}
    r^2 & 1/r^2 & 1/r & 1 \\
     & 1/r^3 & 1/r^2 & 1/r \\
     & & 1/r & 1/r\\
     &  &  & 1 \\
  \end{array}
\right)\ ,\quad
a_\mu^I \sim \mathcal{O}(r, 1/r^2, 1, 1/r) \ ,
\end{equation}
in the basis $(t, r, \theta, \phi)$, where 
$h_{\mu\nu}\equiv \delta g_{\mu\nu}$ and 
$a^I_{\mu}\equiv\delta A_\mu^I$\footnote{Since the background scalar fields~(\ref{GenExt}) are invariant under the Virasoro generators~(\ref{Lam}), one can impose the boundary condition $\delta \chi = 0$. This boundary condition could be relaxed but such an analysis is not needed for our purposes here.}. Moreover, an additional nonlinear boundary condition is imposed to forbid excitations above extremality. The diffeomorphisms and $U(1)^n$-gauge transformations
which preserve the boundary conditions~(\ref{metricbc}) are\footnote{
The boundary conditions for gauge fields are such that only the combined variation $\d\equiv \d_{\xi} + \d_{\Lambda}$ with $\Lambda(\theta,\phi)$ in \eqref{Lam} are asymptotic symmetries.}
\begin{align}
&\xi[\epsilon] = \epsilon(\phi)\p_\phi - r \epsilon'(\phi)\p_r\ ,\no
&\Lambda^I[\epsilon] = -f^I(\theta) \epsilon (\phi)\ ,\label{Lam}
\end{align}
together with  $\xi = \p_t$ and $\Lambda^I=\Lambda^I(t,\theta)$,
which commute with (\ref{Lam}) and (\ref{ASG5}). However, as we will show in section \ref{CC4D}, the latter asymptotic symmetries do not lead to central extensions. Therefore, we will only focus on the Virasoro algebra of the extremal black hole. We take the basis of $\epsilon(\phi)$ as
$\epsilon_n(\phi)=-e^{-in\phi}$ and define 
$\xi_m=\xi[\epsilon_m]$, 
$\Lambda^I_m =\Lambda^I[\epsilon_m]$. Then, the combined generator
$\ell_m\equiv (\xi_m,\Lambda_m)$ 
satisfies the Virasoro algebra with zero central charge as 
\begin{equation}
 i[\ell_m,\ell_n]=(m-n)\ell_{m+n}\ .
\label{Vir4D}
\end{equation}


\subsection{Virasoro algebra in five dimensions}
In five dimensions, there are two boundary conditions to obtain the Virasoro
algebra~\cite{LMP,AOT}. One of them is\footnote{Our previous comments in the 4d case on the boundary conditions for the scalar field, on the additional extremality constraint and on the supplementary asymptotic symmetries are also applicable here.} 
\begin{equation}
h_{\mu\nu} \sim \mathcal{O}
\left(
  \begin{array}{ccccc}
    r^2 & 1/r^2 & 1/r & 1 & r\\
     & 1/r^3 & 1/r^2 & 1/r & 1/r^2\\
     & & 1/r & 1/r & 1/r\\
     &  &  & 1 & 1\\
     &  &  &  & 1/r\\
  \end{array}
\right)\ ,\quad
a_\mu^I \sim \mathcal{O}(r, 1/r^2, 1, 1/r, 1/r) \ ,
\label{bc1}
\end{equation}
in the basis $(t, r, \theta, \phi_1,\phi_2)$.
Another boundary condition is
\begin{equation}
h_{\mu\nu} \sim \mathcal{O}
\left(
  \begin{array}{ccccc}
    r^2 & 1/r^2 & 1/r & r & 1\\
     & 1/r^3 & 1/r^2 & 1/r^2 & 1/r\\
     & & 1/r & 1/r & 1/r\\
     &  &  & 1/r & 1\\
     &  &  &  & 1\\
  \end{array}
\right)\ ,\quad
a_\mu^I \sim \mathcal{O}(r, 1/r^2, 1, 1/r, 1/r) \ .
\label{bc2}
\end{equation}
Then, the asymptotic symmetries are
\begin{align}\label{ASG5}
	\xi_{(i)}[\epsilon] &\equiv \ep (\phi^i) \p_\phi - r\ep'(\phi^i)\p_r \ , \no
	\Lambda_{(i)}^I[\epsilon] &\equiv - f_i^I(\t) \ep(\phi^i) \ , \qquad\qquad (i=1,2) \ ,
\end{align}
where $i=1$ and $i=2$ are for (\ref{bc1}) and (\ref{bc2}) respectively. 
The boundary conditions are not compatible with each other in the sense that there are no consistent boundary
 conditions admitting both set of asymptotic fields or both sets of the Virasoro algebras. 
We take the basis of $\epsilon(\phi^i)$ as
$\epsilon_n(\phi^i)=-e^{-in\phi^i}$ and define 
$\xi_{(i)m}=\xi_{(i)}[\epsilon_m]$, 
$\Lambda^I_{(i)m} =\Lambda^I_{(i)}[\epsilon_m]$. Then, the generator
$\ell_{(i)m}\equiv(\xi_{(i)m},\Lambda_{(i)m})$
satisfies the Virasoro algebra with zero central charge as
\begin{equation}
 i[\ell_{(i)m},\ell_{(i)n}]=(m-n)\ell_{(i)m+n}\ .
\label{Vir5D}
\end{equation}


In the following sections, 
we will calculate the central term of the Virasoro algebra
in~(\ref{Vir4D}) and (\ref{Vir5D}).

\section{Formalism for Conserved Charges}\label{Fomalism}
We need to construct the surface charges which generate the asymptotic symmetries~(\ref{Lam})
and (\ref{ASG5}) to evaluate the central term of the Virasoro algebra
in~(\ref{Vir4D}) and (\ref{Vir5D}). 
In this section, we review the formalism to obtain the conserved charges
for gauge theories following~\cite{BaCo0,BaCo2,Compere,Wald,IyWa,Tachikawa}.

We take the variation of the $D$-form Lagrangian as
\begin{equation}
 \delta\bm{L}(\Phi)=\bm{E}(\Phi)\delta\Phi+d\bm{\Theta}(\delta\Phi,\Phi)\ ,
\label{eq:deltaL0}
\end{equation}
where the $\Phi$ is the generic name of all fields $\Phi = (g_{\mu\nu},A^I_\mu,\chi^A)$.
Then, the equations of motion are given by $\bm{E}(\Phi)=0$. The term 
$\bm{\Theta}(\Phi,\delta\Phi)$ appears as a total divergence and does not
affect the equations of motion. Let $\delta_\eps \Phi$ denotes a general gauge transformation. 
We suppose that the Lagrangian is gauge invariant up to a boundary term 
\begin{equation}
\delta_\eps \bm L (\Phi ) = d \bm M_\eps (\Phi ) \ .\label{defM}
\end{equation}
In the action~(\ref{generalaction}), the gauge symmetries are 
the diffeomorphism and $U(1)^n$ gauge transformations,
$\delta_\epsilon g_{\mu\nu}=\mathcal{L}_\xi g_{\mu\nu}$, 
$\delta_\epsilon A^I_\mu=\mathcal{L}_\xi A^I_\mu+\nabla_\mu \Lambda^I$
and $\delta_\epsilon \chi^A=\mathcal{L}_\xi \chi^A$.
For these gauge transformations, 
the boundary term 
$\bm M_{(\xi,\Lambda)} (\Phi )$ is given by 
\begin{equation}
\bm M_{(\xi,\Lambda)} (\Phi ) = \xi \cdot \bm L(\Phi) + \Lambda \,d \bm{C}_{D-2}(A) \ ,
\end{equation}
where the last contribution appears when the Lagrangian contains a Chern-Simons term of the form 
$\bm{C}_{D}(A) \sim A \wedge F \wedge \dots \wedge F$. 

The quantity $\bm{E}(\Phi)\delta_\eps \Phi$ can be integrated by parts in order to remove the derivatives 
acting on $\eps$ as
\begin{eqnarray}
\bm{E}(\Phi)\delta_\eps \Phi &=& \eps \bm{N}(E(\Phi),\Phi) + d\bm{S}_\eps(E(\Phi),\Phi) \ ,
\nonumber \\
&=& d\bm{S}_\eps(E(\Phi),\Phi) \ . 
\label{defS}
\end{eqnarray}
In the second equality, the Noether identities $\bm{N}(E(\Phi),\Phi) \equiv 0$ were used. 
The Noether current $\bm{S}_\eps(E(\Phi),\Phi)$ associated with the
gauge transformation $\eps$ is vanishing on-shell.
We regard the $\delta$ in (\ref{eq:deltaL0}) as the gauge transformation
$\delta_\epsilon$ and, then, combining 
equations \eqref{eq:deltaL0} and \eqref{defM}, we can express also 
\begin{equation}
\bm{E}(\Phi)\delta_\eps \Phi = -d\bm J_\eps(\Phi)  \ ,\label{secE}
\end{equation}
where we defined the standard covariant phase space Noether current as 
\begin{equation}
 \bm J_\eps(\Phi) = \bm{\Theta}(\delta_\eps\Phi,\Phi)-\bm M_\eps (\Phi)\ .
\label{WaldJ}
\end{equation}
Therefore, the current $\bm{S}_\eps(E(\Phi),\Phi)+\bm J_\eps(\Phi)$ is identically closed and thus exact \cite{Wald2},
 \begin{equation}
 \bm{S}_\eps(E(\Phi),\Phi) = -\bm J_\eps(\Phi) - d\bm{Q}_\eps(\Phi) \ .
\label{defQ}
\end{equation}
Using the properties of the Lie derivative 
$ \mathcal{L}_\xi =\xi\cdot d + d \xi\cdot$ and the on-shell relation $\bm{E}=0$,
we can write the variation of the $\bm M_{(\xi,\Lambda)}$ as
\begin{eqnarray}
\delta \bm M_{(\xi,\Lambda)} (\Phi ) & =& \xi \cdot \delta \bm L(\Phi) + \Lambda \,d \delta \bm{C}_{D-2}(A) \ ,\no
&=& \xi\cdot d \bm{\Theta}(\delta\Phi,\Phi) + \Lambda \,d \delta \bm{C}_{D-2}(A)  \ ,\no
& = & d(- \xi \cdot \bm{\Theta} (\delta\Phi,\Phi) ) + \mathcal L_\xi
 \bm{\Theta}(\delta\Phi,\Phi) + \Lambda \,d \delta \bm{C}_{D-2}(A) \ .
\label{eq:deltaM0}
\end{eqnarray}
Here, following \cite{Tachikawa}, let us define $\bm{\Pi}_{\eps}$ through the equation
\begin{equation}
\delta_{\eps}  \bm{\Theta}(\delta\Phi,\Phi) = \mathcal L_\xi  \bm{\Theta}(\delta\Phi,\Phi) +  \bm{\Pi}_\eps(\delta\Phi,\Phi) \ .\label{eq:39}
\end{equation}
Because the action contains the Chern-Simons terms only for the gauge
fields, we have 
$\bm{\Pi}_{\eps}(\delta\Phi,\Phi) = \bm{\Pi}_{\Lambda}(\delta A,A)$. 
We compute $\delta \delta_\eps \bm{L}$ in two ways: 
we take the gauge transformation $\delta_\epsilon$ of (\ref{defM}) and the variation
 $\delta$ of (\ref{eq:deltaL0}) as
\begin{align}
0 &= \delta \delta_\eps \bm{L}(\Phi ) -  \delta_\eps \delta\bm{L}(\Phi ) \ ,\no
&= d(\delta \bm{M}_\eps (\Phi)- \delta_\eps \bm{\Theta}(\delta_\eps\Phi,\Phi ) ) \ ,\no
&= d(\Lambda d \delta \bm{C}_{D-2}(A) - \bm{\Pi}_{\Lambda}(\delta A,A) ) \ ,
\end{align}
at the last equality, (\ref{eq:deltaM0}) and (\ref{eq:39}) are used.
Therefore, it exists a $(D-2)$-form $\bm{\Sigma}_\Lambda$ such that
\begin{equation}
\Lambda d \delta \bm{C}_{D-2}(A) - \bm{\Pi}_{\Lambda}(\delta A,A) = d\bm{\Sigma}_\Lambda(\delta A,A) \ .
\end{equation}
Using the equation (\ref{eq:deltaM0}) and (\ref{eq:39}), we get
\begin{equation}
\delta \bm M_{(\xi,\Lambda)} (\Phi ) 
 =  d(- \xi \cdot \bm{\Theta} (\delta\Phi,\Phi)
 +\bm{\Sigma}_\Lambda(\delta A,A) ) + \delta_\eps
 \bm{\Theta}(\delta\Phi,\Phi) \ .
\label{eq:deltaM}
\end{equation}
On-shell, we can thus express the variation of the Noether current $\bm{S}_\eps$ as
\begin{eqnarray}
 \delta \bm{S}_\eps(E(\Phi),\Phi) &=& -\delta \bm{\Theta}(\delta_\eps\Phi,\Phi) + \delta \bm M_{(\xi,\Lambda)} (\Phi )
- d \delta\bm{Q}_\eps(\Phi) \ ,\no
&=& \delta_\eps \bm{\Theta}(\delta\Phi,\Phi) - \delta \bm{\Theta}(\delta_\eps\Phi,\Phi)  + d \bm{k}^{cov}_\eps(\delta\Phi,\Phi) \ ,\no
&=& \bm \omega(\delta_\eps \Phi,\delta \Phi)  + d \bm{k}^{cov}_\eps(\delta\Phi,\Phi) \ ,\label{defk}
\end{eqnarray}
where we have defined the surface term 
\begin{equation} 
\bm{k}^{cov}_\eps(\delta\Phi,\Phi) = -\delta \bm{Q}_\eps(\Phi) - \xi \cdot \bm{\Theta} (\delta\Phi,\Phi)
+\bm{\Sigma}_\Lambda(\delta A,A) \ ,
\end{equation}
and the symplectic structure
\begin{equation} 
\bm\omega(\delta_\eps \Phi,\delta \Phi) = \delta_\eps \bm{\Theta}(\delta\Phi,\Phi) 
- \delta \bm{\Theta}(\delta_\eps\Phi,\Phi) \ ,
\end{equation}
which depends on $\eps$ only through the variation of the fields $\delta_\eps \Phi$. 
It follows from \eqref{defk} that $\bm{k}^{cov}_\eps(\delta\Phi)$ is a conserved charge when the equations 
of motion $\bm E(\Phi)=0$, the linearized equations of motion $\delta \bm E(\Phi)=0$ and the symmetry 
conditions $\delta_\eps \Phi = 0$ are satisfied. For asymptotic symmetries, the charges are asymptotically conserved when the equations 
of motion $\bm E(\Phi)=0$ and the linearized equations of motion $\delta \bm E(\Phi)=0$ hold and if the condition 
\begin{equation}
\bm\omega(\delta_\eps \Phi,\delta \Phi)|_{\partial M} = 0 \ ,
\label{asymcon}
\end{equation}
is satisfied. If the surface term is 
integrable,
\begin{equation} 
\bm{k}_\eps(\delta\Phi,\Phi) = \delta \bm{B}_\eps(\Phi) \ ,
\label{inte}
\end{equation}
the equation \eqref{defk} can also be used to define the generator of a
gauge transformation as
\begin{eqnarray}
 Q_\eps[\Phi,\bar\Phi] &=& -\int_C\bm{S}_\eps(E(\Phi),\Phi)+ \int^{\Phi}_{\bar \Phi} \int_{\partial C} \bm{k}_\eps(\delta\Phi,\Phi)+N_\eps[\bar \Phi] \ ,\no
&=& -\int_C\bm{S}_\eps(E(\Phi),\Phi)+ \int_{\partial C} \bm{B}_\eps(\Phi)- \int_{\partial C} \bm{B}_\eps(\bar\Phi)+N_\eps[\bar \Phi]\ ,\label{defCha}
\end{eqnarray}
where $C$ is a Cauchy surface and the integration $\int^{\Phi}_{\bar
\Phi}$ is performed in the phase space of solutions between a reference
solution $\bar \Phi$ and $\Phi$. The boundary term $\int_{\partial C}\bm{B}_\eps(\Phi)$ 
makes $Q_\eps$ differentiable while the background term $\int_{\partial C} \bm{B}_\eps(\bar\Phi)$ may cancel the background divergences. The
term $N_\eps[\bar \Phi]$ is a normalization constant for the reference
solution. Hereafter, we assume the asymptotically conserved
condition~(\ref{asymcon}) and 
the integrability condition~(\ref{inte}).\footnote{
At least, the integrability was
shown for the Virasoro generators on the near horizon geometry of the extreme
Kerr solution~\cite{GHSS}.
}

Now, we make the observation that the definitions on the last line of \eqref{defk} are ambiguous 
by the redefinitions
\begin{equation}
\begin{split}
&\bm\omega(\delta_\eps \Phi,\delta \Phi) \rightarrow \bm\omega(\delta_\eps \Phi,\delta \Phi) 
- d \bm{\mathcal{E}} (\delta_\eps \Phi,\delta \Phi) \ , \\
&\bm{k}^{cov}_\eps(\delta\Phi,\Phi) \rightarrow \bm{k}^{cov}_\eps(\delta\Phi,\Phi) 
+ \bm{\mathcal{E}} (\delta_\eps \Phi,\delta \Phi) \ ,
\end{split}
\label{redef}
\end{equation}
for an arbitrary $\bm{\mathcal{E}} (\delta_\eps \Phi,\delta \Phi)$ anti-symmetric in $\delta_\eps \bm\Phi$ and $\delta \bm\Phi$. This ambiguity generalizes the well-known ambiguity 
in the definition of the pre-symplectic form 
$\bm\Theta(\delta \Phi,\Phi) \rightarrow \bm\Theta(\delta \Phi,\Phi) - d\bm{\mathcal{E}}^\prime(\delta\Phi,\Phi)$ 
which implies \eqref{redef} with 
$\bm{\mathcal{E}}(\delta_\epsilon\Phi,\delta\Phi)=\delta_\epsilon\bm{\mathcal{E}}'(\delta\Phi,\Phi)-\delta\bm{\mathcal{E}}'(\delta_\epsilon\Phi,\Phi)$.
This ambiguity is not relevant for the exact symmetries where $\delta_\eps\Phi = 0$ but has to be fixed 
in the context of the asymptotic symmetries. 

Following the covariant phase space method \cite{Wald,IyWa}, 
one could choose the surface charge $\bm{k}^{cov}_\eps(\delta\Phi,\Phi)$ which does not contain terms proportional 
to $\delta_\eps \Phi$ and its derivatives.

The proposal of \cite{BaCo0,BaCo2,Compere} consists in fixing the surface term $\bm{k}_\eps(\delta\Phi)$ 
by acting on the Noether current $ \bm{S}_\eps(E(\Phi),\Phi)$ with a contracting homotopy $\bm{I}_{\delta \Phi}$. When acting on $D-1$ forms which contain at most second derivatives of the fields, the contracting homotopy $\bm{I}_{\delta \Phi}$ can be written as
\begin{equation}
\bm{I}_{\delta \Phi} = \left( \frac 1 2 \delta \Phi \frac{\partial}{\partial \partial_\mu \Phi} +(\frac{2}{3}\partial_\lambda \Phi - \frac{1}{3}\Phi \partial_\lambda)\frac{\partial}{\partial \partial_\lambda\partial_\mu \Phi}
\right) \frac{\partial}{\partial dx^\mu } \ ,\label{homotopy}
\end{equation}
where the derivative with respect to $dx^\mu$ is defined by
\begin{equation}
 \frac{\partial}{\partial dx^\mu }dx^{\alpha_1}\wedge\cdots
  \wedge dx^{\alpha_p}
=p \,\delta^{[\alpha_1}_\mu dx^{\alpha_2}\wedge\cdots\wedge
  dx^{\alpha_p]}\ .
\end{equation}
This procedure yields a result which depends only on the equations of motion of the Lagrangian. 
The surface term $ \bm{k}_\eps(\delta\Phi)$ can be more easily expressed in terms of the covariant 
phase space expression as
\begin{equation}
 \bm{k}_\eps(\delta\Phi,\Phi) = \bm{k}^{cov}_\eps(\delta\Phi,\Phi)+ \bm{\mathcal{E}}^{hom}(\delta_\eps \Phi,\delta \Phi) \ ,
\label{khom}
\end{equation}
where the supplementary term $\bm{\mathcal{E}}^{hom}(\delta_\eps \Phi,\delta \Phi)$ is given by 
\begin{equation}
 \bm{\mathcal{E}}^{hom}(\delta_\ep \Phi,\delta \Phi) = \frac 1 2 \delta_\ep \Phi \frac{\partial}{\partial \partial_\mu \Phi} 
\frac{\partial}{\partial dx^\mu } \bm \Theta (\delta \Phi,\Phi) \ ,\label{homexpl}
\end{equation}
when $\bm\Theta (\delta \Phi,\Phi)$ contains at most first derivatives
of the fields. 
Here, anti-symmetrization of $\delta \Phi$ and  $\delta_\epsilon \Phi$ factors is understood.

The charge $Q_\eps[\Phi,\bar\Phi]$ generates the asymptotic symmetries
$\eps$ through the covariant Poisson brackets under assumptions about the
integrability, the conservation and the finiteness of the charges as
well as under the condition $\int_{\partial M} \delta \bm{\mathcal{E}}^{hom}(\delta
\Phi,\delta \Phi) = 0$. The algebra of the asymptotic symmetries is the
Poisson bracket algebra of the charges themselves,
\begin{eqnarray}\label{algebra}
\delta_{\tilde \eps} Q_{\eps}[\Phi] \equiv \{ Q_\eps[\Phi,\bar{\Phi}], Q_{\tilde \eps}[\Phi,\bar{\Phi}] \}_{CB} &= & \int_{\partial C} \bm{k}_{\eps}( \delta_{\tilde\eps} \Phi,\Phi) 
\notag\ ,\\
&=& Q_{[\eps, \tilde \eps]}[\Phi,\bar{\Phi}]-N_{[\eps, \tilde \eps]}[\bar \Phi] +
\int_{\partial C} \bm{k}_{\eps}(\delta_{\tilde\eps}\bar{\Phi},\bar{\Phi}) 
\ , 
\end{eqnarray}
where the second line has been obtained from \cite{BaCo2,Compere}. The last term is recognized as the central extension term in the algebra.

\section{Charges for the General Lagrangian in Four dimensions}\label{4Dcharge}
In the previous section, we formally constructed the conserved charges for asymptotic symmetries. 
In this section, we will explicitly calculate the conserved charges of the general
action~(\ref{generalaction}) in four dimensions.

\subsection{General action and equations of motion}
The variation of the four-dimensional Lagrangian~(\ref{generalaction}) is
\begin{equation}
\delta\mathcal{L}=\frac{1}{16\pi}\sqrt{-g}\bigg[
{}^{(g)}E^{\mu\nu}\delta g_{\mu\nu}
+{}^{(A)}E^{\nu}_I\delta A^I_\nu
+{}^{(\chi)}E_C\delta\chi^C\bigg]
+\sqrt{-g}\nabla_\mu X^\mu\ ,
\end{equation}
where ${}^{(g)}E^{\mu\nu}$, ${}^{(A)}E^{\nu}_I$ and
${}^{(\chi)}E_C$ are the equations of motion given by
\begin{align}
&{}^{(g)}E_{\mu\nu}\equiv
-G_{\mu\nu}
-\frac{1}{4}g_{\mu\nu}f_{AB}(\chi)\partial_\rho \chi^A \partial^\rho \chi^B
+\frac{1}{2}f_{AB}(\chi)\partial_\mu \chi^A \partial_\nu \chi^B \nonumber\\
&\hspace{4cm}-\frac{1}{2}g_{\mu\nu}V(\chi)
+\frac{1}{2}k_{IJ}(\chi)\bigg(
F^{I}{}_{\mu\rho} F^{J}{}_\nu{}^\rho-\frac{1}{4}g_{\mu\nu}F^I_{\rho\sigma}F^{J\rho\sigma}
\bigg)=0\ ,
\label{EOMg}\\
&{}^{(A)}E^{\nu}_I\equiv
\nabla_\mu[k_{IJ}(\chi)F^{J\mu\nu}
-\epsilon^{\mu\nu\rho\sigma}h_{IJ}(\chi)F^J_{\rho\sigma}]=0\ ,
\label{EOMA}\\
&{}^{(\chi)}E_C\equiv
-\frac{1}{2}f_{AB,C}(\chi)\partial_\mu \chi^A \partial^\mu \chi^B
+\nabla_\mu(f_{CB}(\chi)\nabla^\mu\chi^B)
-V_{,C}(\chi)\nonumber\\
&\hspace{5.3cm}-\frac{1}{4}k_{IJ,C}(\chi)F^I_{\mu\nu}F^{J\mu\nu}
+\frac{1}{4}h_{IJ,C}(\chi)\epsilon^{\mu\nu\rho\sigma}F^I_{\mu\nu}F^J_{\rho\sigma}=0\ ,
\end{align}
and the total divergence $X^\mu$ is 
\begin{multline}
 X^\mu(\Phi,\delta\Phi)=\frac{1}{16\pi}\big[
(\nabla_\nu h^{\nu\mu} - \nabla^\mu h)\\
\, + \, (-k_{IJ}(\chi)F^{J\mu\nu}
\, + \, h_{IJ}(\chi)\epsilon^{\mu\nu\rho\sigma}F^J_{\rho\sigma})a_\nu^I 
\, - \,f_{AB}(\chi)\nabla^\mu\chi^B \delta\chi^A\big]
\ .
\end{multline}
Here, we define $\Phi=(g_{\mu\nu},A_\mu^I,\chi^A)$, 
$h_{\mu\nu}=\delta g_{\mu\nu}$ and $a_\mu^I=\delta A^I_\mu$.
Then, the variation of 4-form Lagrangian is 
\begin{equation}
 \delta \bm{L}=\bm{E}\delta\Phi + \nabla_\mu X^\mu\bm{\epsilon}
=\bm{E}\delta\Phi + d\ast \bm{X}
\ ,
\label{delL}
\end{equation}
where 
$\bm{E}\delta\Phi\equiv\bm{\epsilon}\left[ {}^{(g)}E^{\mu\nu}\delta g_{\mu\nu}+{}^{(A)}E^{\nu}_I\delta A^I_\nu+{}^{(\chi)}E_C\delta\chi^C \right]/(16\pi)$
and the Hodge dual of $X^\mu$ is defined by
$(\ast\bm{X})_{\alpha\beta\gamma}=X^\mu\epsilon_{\mu\alpha\beta\gamma}$.
From (\ref{eq:deltaL0}) and (\ref{delL}), we can read off the $\bm{\Theta}$ as
\begin{equation}
 \bm{\Theta}(\delta\Phi,\Phi)
=\ast \bm{X}(\delta\Phi,\Phi)\ .
\label{Theta}
\end{equation}
The on-shell vanishing Noether current is given by 
\begin{align}
	S_{(\xi,\Lambda)}^{\m} = \f{1}{16\pi}\left[ 2{}^{(g)}E^\mu_\nu\xi^\n 
	+ {}^{(A)}E_I^\mu (A^I_\rho\xi^\rho+\Lambda^I)\right]\ ,
\end{align}
in the vector form. 
We can rewrite the on-shell vanishing Noether current in the 3-form as
$(\bm{S}_{(\xi,\Lambda)})_{\a\b\g}=S_{(\xi,\Lambda)}^{\m}\epsilon_{\mu\a\b\g}$
and it satisfies $d\bm{S}_{(\xi,\Lambda)}=\bm{E}\delta_{\xi,\Lambda}\Phi$.

\subsection{Current for the diffeomorphism}
The Noether current for the diffeomorphism $\xi$~(\ref{WaldJ}) is
\begin{equation}
 \bm{J}_\xi(\Phi)=\bm{\Theta}(\mathcal{L}_\xi\Phi,\Phi)
-\xi\cdot \bm{L}(\Phi)\ .
\end{equation}
Now, it is convenient to define the vector current $J_\xi^\mu$ by
$(\bm{J}_\xi)_{\alpha_2\alpha_3\alpha_4}\equiv J_\xi^\mu \epsilon_{\mu\alpha_2\alpha_3\alpha_4}$.
Then, the $J_\xi^\mu$ is given by
\begin{equation}
 J_\xi^\mu(\Phi) = X^\mu(\mathcal{L}_\xi\Phi,\Phi)-\xi^\mu L(\Phi)\ ,
\end{equation}
where the $L$ is the Lagrangian which does not include $\sqrt{-g}$, that
is, $\mathcal{L}=\sqrt{-g} L$.
The Lie derivatives for $g_{\mu\nu}$, $A_\mu^I$ and $\chi^A$ are given by
\begin{equation}
\mathcal{L}_\xi g_{\mu\nu} = \nabla_\mu \xi_\nu + \nabla_\nu \xi_\mu\ ,\quad
\mathcal{L}_\xi A_\mu^I
=\xi^\nu F_{\nu\mu}^I+\nabla_\mu (A_\nu^I \xi^\nu)\ , \quad
\mathcal{L}_\xi \chi^A = \xi^\mu \nabla_\mu \chi^A\ .
\end{equation}
We can rewrite $J_\xi^\mu+S^\mu_{(\xi,0)}$ as a total divergence as
\begin{equation}
 J_\xi^\mu(\Phi)+S^\mu_{(\xi,0)}(\Phi)=\nabla_\nu Y^{\mu\nu}_\xi(\Phi) ,
\end{equation}
where $Y^{\mu\nu}_\xi$ is defined by
\begin{equation}
 Y^{\mu\nu}_\xi(\Phi)=\frac{1}{16\pi}\left[\nabla^\nu \xi^\mu-\nabla^\mu \xi^\nu
+(-k_{IJ}(\chi)F^{J\mu\nu}
+h_{IJ}(\chi)\epsilon^{\mu\nu\lambda\sigma}F^J_{\lambda\sigma})
A^I_\rho \xi^\rho \right]\ .
\end{equation}
Therefore, the $\bm{Q}_\xi$ defined by $\bm{J}_\xi+\bm{S}_{(\xi,0)}=-d\bm{Q}_\xi$ is 
\begin{equation}
 \bm{Q}_\xi(\Phi)=-\ast \bm{Y}_\xi(\Phi)\ ,
\label{Qxi}
\end{equation}
where the Hodge dual of $Y^{\mu\nu}_\xi$ is defined by
$(\ast\bm{Y}_\xi)_{\alpha\beta}=(1/2!)Y_\xi^{\mu\nu}\epsilon_{\mu\nu\alpha\beta}$.

\subsection{Current for the $U(1)^n$-gauge transformation}
In the general action~(\ref{generalaction}), there are $U(1)^n$-gauge
symmetries and we can also construct the current of the $U(1)^n$-gauge transformation.
The vector current for the $U(1)^n$-gauge transformation is
\begin{equation}
 J_\Lambda^\mu(\Phi)=
X^\mu(\delta_\Lambda \Phi,\Phi)\ ,
\end{equation}
where the $U(1)^n$-gauge transformations for $g_{\mu\nu}$, $A^I_\mu$ and
$\chi^A$ are 
\begin{equation}
 \delta_\Lambda g_{\mu\nu}=0\ , \quad
\delta_\Lambda A_\mu^I=\partial_\mu \Lambda^I\ ,\quad
\delta_\Lambda \chi^A = 0\ .
\end{equation}
The $J_\Lambda^\m + S^\mu_{(0,\Lambda)}$ can be written as
\begin{equation}
J^\mu_\Lambda(\Phi)+S^\mu_{(0,\Lambda)}(\Phi)=\nabla_\nu Y^{\mu\nu}_\Lambda(\Phi)\ ,
\end{equation}
where $Y^{\mu\nu}_\Lambda$ is defined by
\begin{equation}
 Y^{\mu\nu}_\Lambda(\Phi)=\frac{1}{16\pi}\left(-k_{IJ}(\chi)F^{J\mu\nu}
+h_{IJ}(\chi)\epsilon^{\mu\nu\rho\sigma}F^J_{\rho\sigma}\right) \Lambda^I\ .
\end{equation}
Therefore, the $\bm{Q}_\Lambda$ defined by $\bm{J}_\Lambda+\bm{S}_{(0,\Lambda)}=-d\bm{Q}_\Lambda$ is 
\begin{equation}
 \bm{Q}_\Lambda(\Phi)=-\ast \bm{Y}_\Lambda(\Phi)\ .
\label{QLambda}
\end{equation}

\subsection{Conserved charges}
On-shell, the generator for the diffeomorphism $\xi$ and $U(1)^n$-gauge transformations $\Lambda^I$ is
given by
\begin{equation}
 Q_{\xi,\Lambda}[\Phi,\bar \Phi]
= \int^{\Phi}_{\bar \Phi} \int_{\partial
C}\bm{k}_{\xi,\Lambda}(\delta\Phi,\Phi)+N_\eps[\bar \Phi]\ ,
\label{Q}
\end{equation}
where $\bar{\Phi}$ is the reference solution and $\bm{k}_{\xi,\Lambda}$ is defined by~(\ref{khom}) and can be written as
\begin{equation}
 \bm{k}_{\xi,\Lambda}(\delta\Phi,\Phi)=-\delta \bm{Q}_\xi(\Phi)
-\delta \bm{Q}_\Lambda(\Phi)-\xi\cdot
  \bm{\Theta}(\delta\Phi,\Phi)
+\bm{\mathcal{E}}^{hom}(\delta_{\xi,\Lambda} \Phi,\delta \Phi)\ .
\label{k}
\end{equation}
We can calculate the $\delta \bm{Q}$'s by taking the variation of (\ref{Qxi}) and
(\ref{QLambda}). The expression of $\bm{\mathcal{E}}^{hom}$ can be 
obtained from~(\ref{homexpl}) and (\ref{Theta}). 
We summarize the result as
\begin{equation}
\bm{k}_{\xi,\Lambda}(\delta\Phi,\Phi)
=\bm{k}_\xi^{grav}+\bm{k}_{\xi,\Lambda}^{F}+\bm{k}_{\xi,\Lambda}^{top}+\bm{k}_\xi^{\chi} \ ,
\end{equation}
where
\begin{align}
 &\bm{k}_\xi^{grav}
=
\frac{1}{8\pi} (d^{D-2}x)_{\mu \nu} 
\bigg\{
\xi^\nu\nabla^\mu h
-\xi^\nu\nabla_\sigma h^{\mu\sigma}
+\xi_\sigma\nabla^{\nu}h^{\mu\sigma}
+\frac{1}{2}h\nabla^{\nu} \xi^{\mu}
-h^{\rho\nu}\nabla_\rho\xi^{\mu}\nonumber\\
&\hspace{10cm}
+ \half
h^{\sigma\nu}(\nabla^\mu\xi_\sigma +
\nabla_\sigma\xi^\mu)
\bigg\}\ ,
\label{kgrav}\\
&\bm{k}_{\xi,\Lambda}^{F}=
\frac{1}{16\pi} (d^{D-2}x)_{\mu\nu} 
\bigg[
\bigg\{-k_{IJ,A}(\chi)F^{J\mu\nu}\delta\chi^A
+2k_{IJ}(\chi)h^{\mu\lambda}F^J{}_\lambda{}^{\nu}\nonumber\\
&\hspace{6cm}-k_{IJ}(\chi)\delta F^{J\mu\nu}
-\frac{1}{2}hk_{IJ}(\chi)F^{J\mu\nu} 
\bigg\}(A^I_\rho \xi^\rho+\Lambda^I)\nonumber
\\
&\hspace{7cm}-k_{IJ}(\chi)F^{J\mu\nu}a^I_\rho
 \xi^\rho
-2 \xi^\mu
k_{IJ}(\chi)F^{J\nu\lambda}a^I_\lambda\nonumber\\
&\hspace{8cm}-k_{IJ}(\chi)a^{J\mu}g^{\nu\sigma}
(\mathcal{L}_\xi A^I_\sigma + \partial_\sigma\Lambda^I)\bigg]
\ ,
\label{kF}\\
&\bm{k}_{\xi,\Lambda}^{top}=
\frac{1}{8\pi}
(d^{2}x)_{\mu\nu}
\bigg[
\epsilon^{\mu\nu\lambda\sigma}
\{h_{IJ,A}(\chi)F^J_{\lambda\sigma}\delta\chi^A
+h_{IJ}(\chi)\delta F^J_{\lambda\sigma}
\}(A^I_\rho \xi^\rho+\Lambda^I)\nonumber
\\
&\hspace{5cm}
+\epsilon^{\mu\nu\lambda\sigma}h_{IJ}(\chi)F^J_{\lambda\sigma}a^I_\rho
 \xi^\rho
-2\xi^\nu
h_{IJ}(\chi)\epsilon^{\mu\lambda\rho\sigma}F^J_{\rho\sigma}a^I_\lambda\nonumber\\
&\hspace{7cm}
-2h_{IJ}(\chi)\epsilon^{\mu\nu\rho\sigma}a^J_\rho(\mathcal{L}_\xi A^I_\sigma + \partial_\sigma\Lambda^I)
\bigg]\ ,
\label{kCS}\\
&\bm{k}_\xi^{\chi}=
\frac{1}{8\pi} (d^{D-2}x)_{\mu\nu}  \,
\xi^\nu
f_{AB}(\chi)\nabla^\mu\chi^B \delta\chi^A\ .
\label{kphi}
\end{align}
Here we define $(d^{D-p}x)_{\m_1\dots\m_p} = \frac{1}{p! (D-p)!}\epsilon_{\m_1\dots\m_p\alpha_{p+1}\dots \alpha_{D}}
dx^{\alpha_{p+1}}\wedge \dots \wedge dx^{\alpha_{D}} $, $a_\mu^I= \delta A_\mu^I$, 
$\delta F_{\mu\nu}^I=\partial_\mu a_\nu-\partial_\nu a_\mu$
 and 
$\delta F^{I\mu\nu}=g^{\mu\rho}g^{\nu\sigma}\delta F_{\rho\sigma}^I$.
Now, we are considering the four-dimensional spacetime and substitute $D=4$ into
(\ref{kgrav}), (\ref{kF}) and (\ref{kphi}), but all these
equations except~(\ref{kCS}) are applicable to any $D \geq 2$ as well.

\section{Central Charges for Four-dimensional Extreme Black Holes}\label{CC4D}

Now, we evaluate the central charge for the four-dimensional extremal
black hole. In four dimensions, the near horizon solution~(\ref{GenExt}) 
can be written as 
\begin{equation}
\begin{split}
 &ds^2=\Gamma(\theta)\left[
-r^2dt^2+\frac{dr^2}{r^2}
+\alpha (\t) d\theta^2 \right] + \gamma(\theta)(d\phi+krdt)^2 \ ,\\
 &\chi^A=\chi^A(\theta)\ ,\quad
A^I=f^I(\theta)(d\phi+krdt)\ .
\end{split}
\label{Gen4D}
\end{equation}
We will use this solution as the reference solution $\bar \Phi$. First, we can check that the central extension in the algebra of two asymptotic symmetries generated by $\Lambda_1(\theta,t)$ and $\Lambda_2(\theta,t)$ is zero. Indeed, putting $\xi = 0$, $\delta \chi^ A = 0$, $h_{\mu\nu}=0$, $\Lambda = \Lambda_1(\theta,t)$ and $a_\mu = \partial_\mu \Lambda_2(\theta,t)$, we see that all expressions \eqref{kgrav} to \eqref{kphi} are zero when evaluated on the sphere at infinity $\partial C$.

For the Virasoro generators~(\ref{Lam}), the algebra~(\ref{algebra})
becomes
\begin{equation}
 i\{Q_m,Q_n\}_{CB}=(m-n)Q_{m+n}+
i\int_{\partial C} \bm{k}_{\xi_m,\Lambda_m}((\delta_{\xi_n} 
+ \delta_{\Lambda_n})\bar{\Phi},\bar{\Phi}) - i\, N_{[(\xi_m,\Lambda_m),(\xi_n,\Lambda_n)]}[\bar \Phi]\ ,
\label{alge2}
\end{equation}
where we define $Q_m=Q_{\xi_m,\Lambda_m}[\Phi,\bar{\Phi}]$ and 
$\xi_m$ and $\Lambda^I_m$ are defined under (\ref{Lam}).
The central charge will be read off from the second term on the right-hand side of~(\ref{alge2}).

Because of $(\delta_{\xi_m}+\delta_{\Lambda_m})\chi^A=0$, 
there is no contribution to the central charge 
from $\delta\chi^A$ in  (\ref{kF}), (\ref{kCS}) and (\ref{kphi}).
Thus, the contribution from $\bm{k}^{\chi}$ is zero. The contributions from 
$\bm{k}^{F}$ and $\bm{k}^{top}$ are given by 
\begin{align}
 &i\int_{\partial
  C}\bm{k}^{F}_{\xi,\Lambda}((\delta_{\tilde{\xi}}+\delta_{\tilde{\Lambda}})\bar{\Phi},\bar{\Phi})
\nonumber\\
&\hspace{0.5cm}=-\frac{ik}{16\pi}\int d\theta
d\phi\sqrt{\frac{\alpha(\theta)\gamma(\theta)}{\Gamma(\theta)}}
k_{IJ}(\chi(\theta))f^J(\theta)
\bigg[
\tilde{\epsilon}'(f^I(\theta)\epsilon+\Lambda^I)
-\epsilon'(f^I(\theta)\tilde{\epsilon}+\tilde{\Lambda}^I)
\bigg]
\ ,\label{Fcont}\\
 &i\int_{\partial
  C}\bm{k}^{top}_{\xi,\Lambda}((\delta_{\tilde{\xi}}+\delta_{\tilde{\Lambda}})\bar{\Phi},\bar{\Phi})
\nonumber\\
&\hspace{0.5cm}=-\frac{i}{8\pi}\int d\theta
d\phi \, h_{IJ}(\chi(\theta))
\bigg[
(f^J(\theta)\epsilon+\Lambda^J)_{,\theta}(f^I(\theta)\tilde{\epsilon}+\tilde{\Lambda}^I)'
\no
&\hspace{8cm}
-(f^I(\theta)\tilde{\epsilon}+\tilde{\Lambda}^I)_{,\theta}(f^J(\theta)\epsilon+\Lambda^J)'
\bigg]
\ ,
\label{CScont}
\end{align}
where we put $\xi=\xi[\epsilon]$ and $\tilde{\xi}=\xi[\tilde{\epsilon}]$
and define
$'=\partial_\phi$. 
One finds that the $\bm{k}^F$ and $\bm{k}^{top}$ vanish exactly due
to the relation (\ref{Lam}). The remaining contribution  $\bm{k}^{grav}$ is
\begin{equation}
\begin{split}
&i\int_{\partial
  C}\bm{k}^{grav}_{\xi_m,\Lambda_m}((\delta_{\xi_n}+\delta_{\Lambda_n})\bar{\Phi},\bar{\Phi})\\
&=-\frac{ik}{16\pi}\int d\theta
d\phi\sqrt{\frac{\alpha(\theta)\gamma(\theta)}{\Gamma(\theta)}}
\bigg[\Gamma(\theta)(\epsilon_m'\epsilon_n''
-\epsilon_m''\epsilon_n')
+\gamma(\theta)(\epsilon_m \epsilon_n'
-\epsilon_m' \epsilon_n)
\bigg]\\
&=\frac{k}{4}\delta_{m+n}\left(m^3\int d\theta
\sqrt{\Gamma(\theta)\alpha(\theta)\gamma(\theta)}
+m\int d\theta
\sqrt{\frac{\alpha(\theta)\gamma(\theta)^3}{\Gamma(\theta)}}\right)
\ .
\end{split}
\label{gcont}
\end{equation}
We can read off the central charge from the $m^3$ term in~(\ref{gcont}) as
\begin{equation}
 c=3k \int_0^\pi d\theta
  \sqrt{\Gamma (\theta) \alpha (\theta) \gamma (\theta)}\ .
\label{c4D}
\end{equation}
This is the same result as the one obtained in~\cite{HMNS}. The charges $\bm{k}^{cov}_\eps(\delta\Phi,\Phi)$ defined in the covariant phase space method differ from  $\bm{k}_\eps(\delta\Phi,\Phi)$ by the supplementary contribution $\bm{\mathcal{E}}^{hom}$, see~(\ref{khom}). However, we checked that these charges lead to the same results: the contributions from $\bm{k}^{F},\bm{k}^{top}$ and $\bm{k}^\chi$ are zero and we can obtain the same central charge as~(\ref{c4D}). Therefore, the covariant phase space method~\cite{Wald,IyWa} and the cohomological method~\cite{BaCo0,BaCo2,Compere} give the same central charges.

The term linear in (\ref{gcont}) can be absorbed by an appropriate choice of normalization of $Q_0$. Indeed, if we choose 
\begin{equation}
N_{(\xi_m,\Lambda_m)}  = \delta_{m,0}  \frac{k}{8}\int d\theta
 \sqrt{\frac{\alpha(\theta)\gamma(\theta)^3}{\Gamma(\theta)}}\ ,
 \label{N_0}
\end{equation}
the algebra \eqref{alge2} becomes
\begin{equation}
 i\{Q_m,Q_n\}_{CB}=(m-n)Q_{m+n}+ \frac{c}{12} m^3 \delta_{m+n} \ .
\label{alge3}
\end{equation}

The temperature formula of the left handed dual 
CFT is conjectured in~\cite{HMNS} from the explicit calculation for the Kerr-Newman-AdS
black hole as 
\begin{equation}
 T_L=\frac{1}{2\pi k}\ ,
\end{equation}
and, using the Cardy formula $S_{CFT} = (\pi^2/3) c T_L$, we obtain
the entropy of the dual CFT
\begin{equation}
 S_{CFT} = \frac{\pi}{2}\int_0^\pi d\theta
  \sqrt{\Gamma (\theta) \alpha (\theta) \gamma (\theta)}\ .
\end{equation}
This result agrees with the 
Bekenstein-Hawking entropy of the four-dimensional extremal black hole~(\ref{Sg}).

\section{Extreme Black Holes in Five dimensions}\label{CC5D}
In the previous section, we found that the central charges for the
non-gravitational parts vanish for the fairly general extremal black
holes and we reproduced the Bekenstein-Hawking entropy in four dimensions. 
We will show that this is also the case in five
dimensions. Moreover, as far as the derivation of the charges is
concerned, 
we will keep all formulae general for any odd dimensions 
$D\equiv 2 N + 1$.
We consider the $(2N+1)$-dimensional generalization of the action~(\ref{generalaction}),
\begin{multline}
	S = \f{1}{16\pi}\int d^{2N+1}x \sqrt{-g}\bigg( R - \frac{1}{2}f_{AB}(\chi)
	\partial_\mu \chi^A \partial^\mu \chi^B - V(\chi) \\
	- \frac{1}{4}k_{IJ}(\chi)F^I_{\mu\nu}F^{J\mu\nu} + \frac{1}{2} \tilde F^{\mu\nu}_I F^I_{\mu\nu}\bigg)\ ,
\label{GenAc}
\end{multline}
where
\begin{equation}
\tilde F^{\mu\nu}_I = C_{IJK\dots L}\eps^{\mu\nu\alpha\beta\g\dots \rho\sigma}A^J_\a F^K_{\beta\g}\dots F_{\rho\sigma}^L \ .
\end{equation}
In five dimensions ($N=2$), under some assumptions described in
section \ref{NHofEB}, the near horizon solution of the above theory~(\ref{GenAc})
is given by~(\ref{GenExt})~\cite{KLR}.

\subsection{Conserved charges}
The most of the calculation to obtain the expression of the conserved
charges is the same as the four-dimensional case and 
the gravitational,  the $U(1)$ and the scalar contributions can be read
off directly from \eqref{kgrav}-\eqref{kF}-\eqref{kphi}. 
Thus, what we should consider is only the contribution from the Chern-Simons
term in~(\ref{GenAc}).\footnote{
The conserved charges for the Chern-Simons Lagrangian 
has been already calculated in \cite{BaCo} and in \cite{Rog,Sur,HOT} without the supplementary term $\bm{\mathcal{E}}^{hom}$.}
In the similar fashion as done in section \ref{Fomalism} and \ref{4Dcharge}, 
we can obtain the Chern-Simons contribution for the conserved charge as
\begin{equation} 
\bm{k}^{CS}_{\xi,\Lambda}(\delta\bm A,\bm A) = -\delta \bm{Q}^{CS}_{\xi,\Lambda}(\bm A) - \xi \cdot \bm{\Theta}^{CS} (\delta \bm A,\bm A)
+\bm{\Sigma}_\Lambda(\delta\bm A,\bm A)
+\bm{\mathcal{E}}^{hom}(\mathcal L_\xi \bm A + d \Lambda, \delta \bm A)
\ ,
\label{kCShom0}
\end{equation}
where 
\begin{align}
&\bm Q^{CS}_{\xi,\Lambda}(\bm A) = - \frac{N}{16\pi}
 (d^{D-2}x)_{\mu\nu}\left[  \tilde F^{\mu\nu}_I (A^I_\rho \xi^\rho +
		     \Lambda^I)\right]\ ,\\
&\bm \Theta^{CS}(\delta\bm A,\bm A) = \frac{N}{16\pi} (d^{D-1}x)_\mu \left[ \tilde F^{\mu\nu}_I \delta A_\nu^I \right] \ ,\\
&\bm{\Sigma}_\Lambda(\delta\bm A,\bm A) 
=\frac{N}{16\pi}(d^{D-2}x)_{\mu\nu}C_{IJK\cdots L}
\Lambda^I\epsilon^{\mu\nu\alpha_3\alpha_4\cdots\alpha_{D-1}\alpha_D}
a_{\alpha_3}^J F_{\alpha_4\alpha_5}^K \cdots
 F_{\alpha_{D-1}\alpha_D}^L \ ,\\
&\bm{\mathcal{E}}^{hom}(\mathcal L_\xi \bm A + d \Lambda, \delta \bm A)
 = -\frac{N(N-1)}{8\pi } (d^{D-2}x)_{\mu\nu} (\mathcal L_\xi
 A^I_{\beta} + \partial_\beta  \Lambda^I)\nonumber \\
& \hspace{8cm}
\times C_{IJK\dots L} \eps^{\mu\nu \alpha\beta\gamma \dots
 \rho\sigma}\delta A^J_\alpha A^K_\gamma \dots F_{\rho\sigma}^L \ .
\end{align}
A shorter route to find the expression for the Chern-Simons contribution consists in first writing the Noether current,
\begin{equation*}
\bm{S}_\eps^{CS}(\bm{A} ) = \frac{1}{16\pi} (d^{D-1}x)_\mu  \frac{N+1}{2}C_{IJ\dots K}\eps^{\mu\alpha\beta \dots \rho\sigma}F^J_{\alpha\beta} \dots F_{\rho\sigma}^K
(A_\rho^I \xi^\rho + \Lambda^I) \ .
\end{equation*}
Since the current depends at most on the first derivatives of $\bm{A}$, only the first term in \eqref{homotopy} contributes, and we get as a result
\begin{equation}
 \bm{k}^{CS}_{\xi,\Lambda}(\delta\bm A,\bm A)
=\frac{N(N+1)}{16 \pi} (d^{D-2}x)_{\mu\nu}  ( C_{IJ\dots K} \eps^{\mu\nu \gamma \alpha \beta \dots \rho\sigma}\delta A^J_\gamma  F^K_{\alpha\beta}\dots  F^L_{\rho\sigma} )(A_\rho^I \xi^\rho + \Lambda^I) \ .\label{kCShom}
\end{equation}
We have checked that the expressions \eqref{kCShom0} and \eqref{kCShom} only differ by a total derivative and that the expression \eqref{kCShom} is identical to the one found in \cite{BaCo}.

\subsection{Central charge}
Let us calculate the central charge for the near horizon metric of the
five-dimensional extremal black holes
(\ref{GenExt}). 
For each set of the boundary conditions~(\ref{bc1}) and (\ref{bc2}),
there are two asymptotic symmetries given in~(\ref{ASG5}).
For each of these sets of asymptotic symmetries, the contribution to
the central term  from
the scalar fields is zero because of $(\d_{\xi_i} +
\d_{\Lambda_i})\chi=0$.
The contributions from $\bm{k}^{F}$ and $\bm{k}^{CS}$ are 
\begin{align}
	&i\int_{\p C}\bm{k}^{F}_{\xi_{(i)}, \Lambda_{(i)}}
((\delta_{\ti\xi_{(i)}} + \delta_{\ti\Lambda_{(i)}})\bar{\Phi},\bar{\Phi})\no
	&= -\frac{i}{16\pi}\int d\theta d\phi^1 d\phi^2 
		\s{\f{\a(\t)\gamma(\theta)}{\G(\t)}}\,
		k_{IJ}(\chi) \sum_j k^j f_j^J(\t)\bigg[ 
        \ti\ep'(f^I_i\epsilon+\Lambda^I_{(i)})-\ep'(f^I_i\ti\epsilon+\ti\Lambda^I_{(i)})
	\bigg] \ , \\
	&i\int_{\p C}\bm{k}^{CS}_{\xi_{(i)}, \Lambda_{(i)}}
((\delta_{\ti\xi_{(i)}} + \delta_{\ti\Lambda_{(i)}})\bar{\Phi},\bar{\Phi})\no
	&= -\frac{3i}{8\pi}\int d\theta d\phi^1 d\phi^2 
		C_{IJK} f^I_{j,\theta}\bigg[ 
	(f^J_i\epsilon+\Lambda^J_{(i)})(f^K_i\ti\epsilon+\ti\Lambda^K_{(i)})'
	- (f^J_i\epsilon+\Lambda^J_{(i)})' (f^K_i\ti\epsilon+\ti\Lambda^K_{(i)})\bigg] \ ,
\label{5dCS}
\end{align}
where we put $\xi_{(i)}=\xi_{(i)}[\epsilon]$, $\tilde\xi_{(i)}=\xi_{(i)}[\tilde\epsilon]$,
$\gamma=\text{det}(\gamma_{ij})$, $'=d/d\phi^i$ and, in (\ref{5dCS}),
$j\neq i$. 
Substituting the explicit form of $\Lambda^I_{(i)}$ in~(\ref{ASG5}), 
we can find that the contributions from $\bm{k}^{F}$ and $\bm{k}^{CS}$
vanish exactly.
The contribution from $\bm{k}^{grav}$ is
\begin{align}
 &i\int_{\p C}\bm{k}^{grav}_{\xi_{(i)m}, \Lambda_{(i)m}}
((\delta_{\xi_{(i)n}} + \delta_{\Lambda_{(i)n}})\bar{\Phi},\bar{\Phi})\no
&= - \frac{i}{16\pi}\int d\theta d\phi^1 d\phi^2 
\s{\f{\a(\t)\gamma(\theta)}{\G(\t)}}
\bigg[ k^i\G(\t)(\ep_m'\ep_n'' - \ep_m''\ep_n')
+ \sum_{j}k^j\g_{ij}(\t)(\ep_m \ep_n' - \ep_m' \ep_n)
\bigg] \ , \no
&=\frac{(2\pi)^2}{8\pi}\delta_{m+n}\bigg[
m^3 k^i\int d\theta \sqrt{\Gamma(\theta)\alpha(\theta)\gamma(\theta)} 
+ m\int d\theta
 \sqrt{\frac{\alpha(\theta)\gamma(\theta)}{\Gamma(\theta)}} 
\sum_{j}k^j\gamma_{ij}(\theta)
\bigg]\ .
\label{5dcgrav}
\end{align}
From the $m^3$ term in~(\ref{5dcgrav}), 
the central charges are found to be 
\begin{align}\label{GenCCHigher}
 c_i =   6 \pi k^i \int_0^\pi d\t \s{\G (\t) \a (\t) \g (\t)} \quad
 \text{for }\,i=1,2\ .
\end{align}
Even if we put $\bm{\mathcal{E}}^{hom}=0$, we can get the same central
charge. 
So the covariant phase space methods~\cite{Wald,IyWa} and the
cohomological methods~\cite{BaCo0,BaCo2,Compere} give the same results even in five dimensions.

The temperature formulae of dual CFTs are conjectured in~\cite{CCLP} starting from the higher-dimensional Kerr-AdS black holes as
\begin{equation}
 T_i=\frac{1}{2\pi k^i}\quad
 \text{for }\,i=1,2\ .
\end{equation}
Thus, using the Cardy formula, we can obtain
the entropy of the dual CFTs as
\begin{equation}
 S_{CFT} = \frac{\pi^2}{3}c_1 T_1
=\frac{\pi^2}{3}c_2 T_2
=\pi^2\int_0^\pi d\theta
  \sqrt{\Gamma (\theta) \alpha (\theta) \gamma (\theta)}\ .
\end{equation}
The two boundary conditions~(\ref{bc1}) and (\ref{bc2}) give the same entropy.
This result coincides with the 
Bekenstein-Hawking entropy of the five-dimensional extremal black hole~(\ref{Sg}).
\section{Conclusion}\label{Conc}

Any extremal black hole in generic $4d$ Einstein-Maxwell-scalar theory
with topological terms and $5d$ Einstein-Maxwell-Chern-Simons-scalar
theory has a near-horizon geometry whose asymptotic symmetries contain
one (in $4d$) or two (in $5d$) centrally-extended Virasoro
algebra(s). We checked that only the Einstein Lagrangian contributes to
the value of the central charge, and therefore, assuming the conjectured
temperature, that the Bekenstein-Hawking entropy of any extremal black
hole is correctly reproduced. 
These results support the extreme black hole/CFT duality suggested
in~\cite{HMNS}.

The central charges have been computed
using both cohomological and covariant phase space methods and have
shown to agree. Our results are expected to hold in any dimension,
because the expressions for the charges and the near-horizon metric are straightforward generalizations of the four and five-dimensional cases.

In the derivation of the central charge, we used the general action (\ref{generalaction}) and 
the near horizon extremal metric (\ref{GenExt}) which can be obtained from (\ref{generalaction}) 
under very mild assumptions. In particular, the result holds for any topology of the horizon except $T^{D-2}$.
Therefore, our result is applicable to five-dimensional black holes with non-trivial topology
such as the black rings \cite{EmRe,PoSe} as long as the horizon is simply connected. Extremal black saturns and di-rings solutions are not known but are conjectured to exist (see e.g. \cite{ElFi,IgMi,Evslin:2007fv,Izumi,ElRo}). In the case of the black holes with disconnected horizons, including the extremal black saturns and di-rings, we could apply the extreme black hole/CFT correspondence to each horizon. Then the Bekenstein-Hawking entropy would be reproduced as the sum of the entropies of dual CFTs.

\vspace{1cm}
\centerline{\bf Acknowledgements}

We are grateful to T. Azeyanagi, G. Barnich, S. de Buyl, S. Detournay, D. Marolf, T. Hartman, N. Ogawa, A. Strominger, Y. Tachikawa, T. Takayanagi and S. Terashima 
for valuable discussions, and
especially to T. Hartman and A. Strominger for collaboration at an earlier stage.
The work of KM and TN are supported by JSPS Grant-in-Aid for Scientific Research
No.\,19$\cdot$3715 and No.\,19$\cdot$3589 respectively. The work of G.C. is supported in part by the US National Science Foundation under Grant No.~PHY05-55669, and by funds from the University of California.



\end{document}